\documentstyle[pre,preprint,aps,amsmath,epsfig,amssymb]{revtex}

\begin{document} 
\draft
\title{A cellular automaton traffic flow model between the 
Fukui-Ishibashi and Nagel-Schreckenberg models}
\author{Lei Wang$^1$, Bing-Hong Wang$^{2,3,4}$ and Bambi Hu$^{1,5}$}
\address{
$^1$ Physics Department and Centre for Nonlinear Studies, \\
     Hong Kong Baptist University, Hong Kong, China \\
 $^2$ CCAST (World Laboratory), P.O.Box 8730, Beijing 100080, 
China \\ 
$^3$ Center of Nonlinear Science and Department of Modern 
Physics,\\
     University of Science and Technology of China, Hefei 230026\\
$^4$ Department of Physics, The Chinese University of Hong Kong,\\
     Shatin, New Territories, Hong Kong \\
 $^5$ Department of Physics, University of Houston, Houston, 
TX 77204, USA}
\maketitle
%\vspace*{0.15 true in}
\begin{abstract} 
We propose and 
study a new one-dimensional traffic flow cellular automaton (CA)
model of  high speed vehicles with the 
Fukui-Ishibashi-type acceleration
for all cars and the Nagel-Schreckenberg-type (NS) stochastic delay 
only for the cars following the trail of the car ahead.  
The main difference in the 
delay scenario between the new model and the 
NS model is that a car with spacing
ahead longer than the 
velocity limit $M$ may not be delayed in the new model.
By using a car-oriented mean field theory, we derive the fundamental
diagrams of the average speed as the function of car density
 analytically. Our theoretical results are in excellent 
agreement with numerical simulations.
\end{abstract} 


\allowdisplaybreaks 

\section{INTRODUCTION}

Traffic flow cellular automaton(CA) models 
have attracted much interest recently. 
Compared with other dynamical approaches, e.g. the fluid dynamical 
approach, to this class of problems, 
CA models are conceptually simpler
and can be easily implemented on computers for numerical 
investigations [1-4].

Two popular one-dimensional (1D) 
traffic flow models are the Fukui-Ishibashi(FI) model [5] and 
the Nagel-Schreckenberg (NS) model [6]. An exact 
car-oriented mean-field (COMF) theory 
has been developed for the FI model with 
arbitrary limit on the maximum speed $v_{\mbox{max}}$, car density 
$\rho$ and delay probability $f$ [7,8].
However, for the 
NS model with high speed vehicles ($v_{\mbox{max}}>1$) and 
stochastic delay, there is still no established 
exact analytical theory up to now [9,10].

The acceleration and stochastic delay rules of the 
NS model lead to complications  
in the time evolution of the flow and hence 
it is very difficult for exact analytical studies.   
In order to understand how these rules affect the evolution and 
the corresponding asymptotic state, 
we study a new 1D traffic flow CA model in which only
the cars following the trail of the car ahead may be delayed.  

The plan of the present paper is as follows.  
The definition of the model and the evolution equations 
for the inter-car spacings are given in Sec. II. In Sec. III 
some observations are made to describe the
steady state of the system. We present the fundamental diagrams for
the low car density case with arbitrary vehicle speed limit and the 
high density case with vehicle speeds limited to 1 and 2 in Sec. IV.
Excellent agreements between numerical simulations and theoretical 
results are shown in Sec. V together with a discussion on our 
results in connection to the FI and NS models. 

\section{THE MODEL}

Let $N$ be the total number of cars on a 1D 
road of length $L$. 
The density of cars is $\rho=N/L$. Let $C_n(t)$ 
be the number of empty sites in 
front of the $n$th car at time $t$, and $v_n(t)$ 
be the number of sites that the
$n$th car moves during the time $t$ step. 

The new model adopts the following acceleration rule [5]: 

\noindent {\em Step 1}: $v'_n(t)=\min(C_n(t),M)$

\noindent We call the $n$th car ``a car that follows the trail 
of the car ahead" if 
$v'_n(t)=C_n(t)$. It means that the $n$th car may become 
the neighbor of the car ahead if the car in front stops. 
Stochastic delay is introduced in such a way that 
all the cars which follow the trail of 
their cars ahead have a probability $f$ to move forward 
one site less than it is allowed by {\em Step 1}, i.e., we have 

\noindent {\em Step 2}: $v_n(t)=v'_n(t)-1$ with the 
probability $f$,  if $v'_n(t)=C_n(t)$ and $v'_n(t)>0$, 

\noindent and 

\noindent {\em Step 3}: The $n$th car moves $v_n(t)$ sites ahead.

The number of empty sites in front of the $n$th car
 at time $t+1$ can be written as
\begin{equation} \label{ww31} 
C_n(t+1)=C_n(t)+v_{n+1}(t)-v_n(t) 
\end{equation}
For the new model with maximum car velocity 
$v_{\mbox{max}}=M$ and a stochastic delay
probability $f$, the velocity of the $n$th car at time step $t$ as a 
function of the inter-car
spacing $C_n(t)$ can be written as
\begin{equation} \label{ww32}v_n(t)=F_M(f,C_n(t)) \end{equation}
where
\begin{equation} \label{ww33}
     F_M(f,C)=\left\{  \begin{array}{ll}
				  \quad  \;\:  
	M               &   \mbox{if}\; C>M \\
				       \left. \begin{array}{clc}
		 C-1 \quad    &\mbox{with probability} &f \\
	  C \quad  &\mbox{with 
probability}&1-f                                                    
	\end{array} \right\}&     \mbox{if}\; 0<C\leq M \\
       \quad \;\:    0              &     \mbox{if}\; C=0
     \end{array}
	   \right.
\end{equation}

\section{INTER-CAR SPACINGS IN THE STEADY STATES}

From Eqs.(\ref{ww31})-(\ref{ww33}), we can derive the 
properties of the inter-car spacings in the steady states.  
Given $C_n(t)\leq M+1$, it follows that 
$C_n(t)-F_M(C_n(t))\leq 1$, 
  and from $F_M(C_{n+1}(t))\leq M$, we obtain 
   $C_n(t+1)=C_n(t)-F_M(C_n(t))+F_M(C_{n+1}(t))\leq M+1$.
Therefore, if an inter-car spacing is not larger 
than $M+1$, it will not be larger than $M+1$ as the system 
evolves.  

Given $C_n(t)\geq M+1$, it follows that 
$F_M(C_n(t))=M$, and from $F_M(C_{n+1}(t))\leq M$ we have 
$C_n(t+1)=C_n(t)-F_M(C_n(t))+F_M(C_{n+1}(t))\leq C_n(t)$.  
Therefore, inter-car spacings which are larger than 
or equal to $M+1$ will never increase, 
i.e., if $C_n(t)\geq M+1$, then $C_{n+1}(t)\leq C_n(t)$.

It is useful to define the long and short inter-car spacings via 
their comparison with the maximum car speed $M$.
An inter-car spacing is called a long spacing if 
it is longer than $M+1$, i.e., 
$C_n(t)>M+1$.
An inter-car spacing is called a short spacing if it 
is shorter than $M+1$, i.e.
$C_n(t)<M+1$.
Based on the above definitions, we can define 
the excessive length of a long spacing $L_n(t)$ and 
the deficient length of a short spacing $S_n(t)$ as  
$$L_n(t)=\max(C_n(t)-(M+1),0)$$
and 
$$S_n(t)=\max((M+1)-C_n(t),0).$$ 
It follows that 
the sum of the excessive lengths of all long spacings $L(t)$ 
and the sum of deficient lengths of 
all short spacings $S(t)$ are given respectively by 
$$L(t)=\sum\limits_n L_n(t), \;\;\;  S(t)=\sum\limits_n S_n(t)$$ 

From these definitions, it can be proven readily that 
\begin{align}
\label{const} L(t)-S(t)&=\sum\limits_n [C_n(t)-(M+1)]  \nonumber \\ 
		       &=L-(M+2)  \nonumber \\ 
		       &=\mbox{constant}.
\end{align}  
From these properties of the 
inter-car spacings, we have  
$\label{Lndecrease} L_n(t+1)\leq L_n(t)$. 
Hence,  
\begin{equation}\label{Ldecrease} 
L(t+1)\leq L(t). 
\end{equation} 
From Eqs.(\ref{const}) and (\ref{Ldecrease}), we have 
$S(t+1)\leq S(t)$.   
Therefore $L$ and $S$ will never increase as the system 
evolves.  If  
one of the $L_n$ decreases, then $L$ and $S$ will have to decrease.

Next we look into the question of whether long and short spacings
may co-exist in the asymptotic steady state. 
Let $N_i(t)$ be the number of inter-car spacings with length 
$i$ at time $t$.  The probability of finding such a spacing at time 
$t$ is $P_i(t)=N_i(t)/N$. 
Hereafter, $P_i(t)$ is denoted by $P_i$ for simplicity, except 
specified otherwise.
Suppose that long and short spacings co-exist.  
Consider a long spacing, if the car ahead moves forward by $m-1$ 
sites, then the spacing will decrease by 1.  
The probability for this to happen is $(1-f)P_{M-1}+fP_M$. 
For the same reason, the probability for the spacing 
to be shortened by 2 is  
$(1-f)P_{M-2}+fP_{M-1}$, and the probability for the 
spacing to be shortened by 3 is $(1-f)P_{M-3}+fP_{M-2}$, and so on. 
The probability for the spacing to be 
shortened by $M-1$ is $(1-f)P_1+fP_2$, and 
the probability for the spacing to be 
shortened by $M$ is $P_0+fP_1$. 
On average, a long spacing will be shortened by
    \begin{eqnarray}\label{longdecrese} 
     MP_0+\sum\limits_{i=1}^{M-1}{[(1-f)(M-i)+f(M-i+1)]P_i}+fP_M 
\nonumber \\ 
     =MP_0+\sum\limits_{i=1}^{M-1}{(M-i+f)P_i}+fP_M  
     \end{eqnarray} 
in one time step.  The shortened length is positive, 
unless $P_0=P_1=P_2=......P_{M-1}=P_M=0$, i.e., $S=0$.
Therefore, in the asymptotic steady state of the system, 
$L$ and $S$ will no longer change and at least one of them 
becomes zero.  Hence, it is not possible for long and short 
spacings to co-exist in the asymptotic steady state.

\section {ANALYTICAL SOLUTION OF ASYMPTOTIC VELOCITY}

For the low car density case ($\rho < 1/(M+2)$), it is 
apparent that
in the asymptotic steady state, $L>0$ and $S=0$.  Hence
\begin{equation} \label{low} 
C_n(t)\geq M+1,\qquad \forall n. 
\end{equation}
In this case, stochastic delay will no longer occur and 
all the cars will move forward with the 
maximum speed $M$. The average car speed of traffic flow is 
\begin{equation} 
<V(t\rightarrow \infty )>=M. 
\end{equation}

For the high car density case ($\rho > 1/(M+2)$),  it is 
apparent that
in the asymptotic steady state $S>0$ and $L=0$.  Hence
\begin{equation}
C_n(t)\leq M+1 ,\qquad \forall n. 
\end{equation}
The length of every inter-car spacing cannot be larger than $M+1$. 
Therefore, the average speed of traffic flow in the 
asymptotic steady state is
   \begin{align}  
     <V(t\rightarrow \infty )>
       &=\sum\limits_{i=1} ^M {P_i[i(1-f)+(i-1)f]}+MP_{M+1} 
	\nonumber \\
       &=\sum\limits_{i=1} ^M {P_i(i-f)}+MP_{M+1} \label{ww3VP}
   \end{align}

\subsection{$v_{\mbox{max}}=M=1$}
In this case, the high density case refers to $\rho \geq 1/3$, and 
hence $P_n=0,\quad \forall n\geq 3$.  It implies that 
only $P_0$, $P_1$, $P_2$ are non-zero.  
To obtain the non-vanishing $P_j$, 
we introduce $N_{i \rightarrow j}$ to 
describe the number of inter-car spacings with 
a change in length from $i$ at time $t$ to $j$ at time $t+1$. 
The probability of finding an inter-car spacing with length $i$ 
at time $t$ and length $j$ at time $t+1$ is
\begin{equation}
W_{i\rightarrow j}(t) \equiv N_{i\rightarrow j}(t) /N. 
\end{equation}
From Eqs.(\ref{ww31})-(\ref{ww33}), 
we can write down all the non-zero $W_{i\rightarrow j}$ as 
\begin{align*} W_{0 \rightarrow 1}&=P_0[(1-f)P_1+P_2] \\
	       W_{0 \rightarrow 2}&=0 \\
	       W_{1 \rightarrow 0}&=P_1[(1-f)P_0+f(1-f)P_1] \\
	       W_{1 \rightarrow 2}&=P_1[f(1-f)P_1+fP_2]\\
	       W_{2 \rightarrow 0}&=0 \\
	       W_{2 \rightarrow 1}&=P_2(P_0+fP_1)
\end{align*} 
For $|i-j|\geq 3$, $W_{i \rightarrow j}=0$. 

When the system approaches its asymptotic steady state, 
all the $P_j$ cease to change.
So the following detailed balance condition for the 
steady state holds:
      \begin{equation}\label{de} \sum\limits_{i \not= m}W_{i 
\rightarrow m}=
	  \sum\limits_{i \not= m}W_{m \rightarrow i}, 
	  \qquad \forall m. 
	  \end{equation}
When $m$ equals $0$,$1$, and $2$, three detailed balance 
equations can be written down as:
      \begin{align*} 
      W_{0 \rightarrow 1}+W_{0\rightarrow 2}&=
      W_{1 \rightarrow 0}+W_{2 \rightarrow 0}\\
      W_{1 \rightarrow 0}+W_{1 \rightarrow 2}&=W_{0 \rightarrow 
1}+W_{2 \rightarrow 1}\\
      W_{2 \rightarrow 0}+W_{2 \rightarrow 1}&=W_{0 \rightarrow 
2}+W_{1 \rightarrow 2}
      \end{align*} 
Substituting the expressions of $W_{i \rightarrow j}$ into the 
above three equations, we obtain
      \begin{equation}\label{m1eq} P_0P_2=f(1-f)P_1^2 .
       \end{equation}
Normalization requires that $P_0$, $P_1$, and $P_2$ satisfy 
the equations
      \begin{equation}\label{ww3m1g1} 
      \sum\limits_i P_i=P_0+P_1+P_2=1, \end{equation}
and 
      \begin{equation}\label{ww3m1g2} \sum\limits_i 
      iP_i=P_1+2P_2=\bar C=1/\rho -1. \end{equation}

From Eqs.(\ref{m1eq}),(\ref{ww3m1g1}), and (\ref{ww3m1g2}), 
we obtain a quadratic equation for $P_0$:
      \begin{equation}\label{ww3m1eq} 
	(2f-1)^2P_0^2+[(2f-1)^2(\bar C -2)+1]P_0-f(1-f)
	(\bar C -2)^2=0 , 
      \end{equation}
with its root given by 
      \begin{equation}\label{ww3m1p} 
	  P_{0}=\frac{-\lbrack (2f-1)^{2}(\bar C -2)+1\rbrack 
		 +\sqrt{(2f-1)^{2}(\bar C -2)\bar C +1}}
		 {2(2f-1)^{2}} .
      \end{equation}
Hence, the asymptotic average speed of traffic flow is 
       \begin{align}  
	<V(t\rightarrow \infty )>
&=\frac{\bar C +\frac{1}{2f-1}(\sqrt{(2f-1)^{2}(\bar C -2)\bar C 
+1}-1)}{2}\nonumber\\
&=\frac12 \left(
-1+\frac1\rho+\frac{-1+\sqrt{1+\frac{(2f-1)^2(\rho-1)(3\rho-1)}
{\rho^2}}}{2f-1}
   \right)  \label{ww3m1VP}
      \end{align}
For $f=1/2$, which is an removable singular point,
      \begin{equation}<V(t\rightarrow \infty )>=\bar C /2 =(1/ 
\rho-1)/2 . 
      \end{equation}
Equations (\ref{ww3m1VP}) and (19) give the asymptotic 
$<V(t\rightarrow \infty)>$ as a function of $f$ 
and $\rho$ in the high                                                 
density case with $M=1$.

\subsection{$v_{\mbox{max}}=M=2$}
In this case, $\rho \geq 1/4$, 
hence $P_n=0,\quad \forall n\geq 4$.  It implies that 
only $P_0$, $P_1$, $P_2$ and $P_3$ are non-zero. 
From Eqs.(\ref{ww31})-(\ref{ww33}), 
we can write down the non-zero $W_{i\rightarrow j}$ as:
       \begin{align*} 
	       W_{0 \rightarrow 1}&=P_0[(1-f)P_1+fP_2] \\
	       W_{0 \rightarrow 2}&=P_0[(1-f)P_2+P_3] \\     
	       W_{0 \rightarrow 3}&=0 \\
	       W_{1 \rightarrow 0}&=P_1[(1-f)P_0+f(1-f)P_1] \\
	       W_{1 \rightarrow 2}&=P_1 \{ 
f(1-f)P_1+[f^2+(1-f)^2]P_2+(1-f)P_3 \} \\
	       W_{1 \rightarrow 3}&=P_1[(1-f)P_2+fP_3] \\
	       W_{2 \rightarrow 0}&=P_2[(1-f)P_0+f(1-f)P_1] \\
	       W_{2 \rightarrow 1}&=P_2 \{ 
fP_0+[f^2+(1-f)^2]P_1+(1-f)P_2 \} \\
	       W_{2 \rightarrow 3}&=P_2[f(1-f)P_2+fP_3] \\
	       W_{3 \rightarrow 0}&=0 \\
	       W_{3 \rightarrow 1}&=P_3(P_0+fP_1) \\
	       W_{3 \rightarrow 2}&=P_3[(1-f)P_1+fP_2] 
       \end{align*} 
Substituting the above expressions into the detailed balance
condition in Eq.(\ref{de}), we obtain the following set of four 
equations:
      \begin{equation}\label{ww3m20} 
	 fP_0P_2+P_0P_3-f(1-f)P_1^2-f(1-f)P_1P_2=0
      \end{equation}
      \begin{equation}\label{ww3m21}
 2fP_0P_2+P_0P_3-2f(1-f)P_1^2-f(1-f)P_1P_2-(1-f)P_1P_3+f(1-f)P_2^2=0
      \end{equation}
      \begin{equation}\label{ww3m22}
	 fP_0P_2-P_0P_3-f(1-f)P_1^2+f(1-f)P_1P_2-2(1-f)P_1P_3
	 +2f(1-f)P_2^2=0       
      \end{equation}  
      \begin{equation}\label{ww3m23}           
	 P_0P_3-f(1-f)P_1P_2+(1-f)P_1P_3-f(1-f)P_2^2=0
      \end{equation}  
Noting that only two of them, e.g. 
Eqs. (\ref{ww3m20}) and (\ref{ww3m23}), are independent.
Combining Eqs.(\ref{ww3m20}) and (\ref{ww3m23}) 
with the normalization conditions
      \begin{equation}\label{ww3m2g1} 
      \sum\limits_i P_i=P_0+P_1+P_2+P_3=1 \end{equation}
and       
      \begin{equation}\label{ww3m2g2} \sum\limits_i iP_i
      =P_1+2P_2+3P_3=\bar C=1/\rho -1 ,
      \end{equation}
we can solve for $P_0$, $P_1$, $P_2$, $P_3$ and 
obtain the asymptotic traffic flow velocity
for $M=2$.

\vspace*{0.15 true in}

\section{DISCUSSION}

In order to compare with the analytic results, we carried out 
numerical 
simulations on 1D chain with $1000$ cars. The length of 
the chain was adjusted so as to give the desired car density.  
Periodic boundary condition was imposed.  
The first $20000$ time steps 
were excluded from the averaging procedure so as 
to remove the transient behavior. The averages were taken over 
the next $80000$ time steps. Figures 1 and 2 show the comparison 
between results obtained from numerical simulations 
and our mean field 
theory for the cases of $M=1$ and $M=2$ over the entire 
range of density $\rho$.
The curves are the theoretical results while the symbols represent 
results of numerical simulations. The curves from the top down 
along the velocity axis correspond 
to different values of $f$ ranging 
from 0 to 1. Excellent agreement between simulations and our theory 
is found.

From the fundamental diagrams of the new model, it is noted that when 
the car density is low enough ($\rho \leq 1/(M+2)$),
all the inter-car spacings will not be shorter than $M+1$,
and all the cars will not be delayed, leading to traffic flow in
 its maximum velocity ($V=M$). This situation is more realistic in 
that 
in real traffic, no driver would like to slow down his car 
when it is far away from the car ahead.
In the high density case, the stochastic delay in our new model
represents better safety than that of the FI model, and leads to
much higher asymptotic average velocity of traffic flow than that
in the NS model.

In summary, we introduced a new model with stochastic delays 
for cars following the trail of the car ahead.
Its evolution and fundamental diagram are 
quite different from the NS and FI models, 
even in the simplest case of $M=1$. 
We studied the evolution of the inter-car spacings and obtained its 
fundamental diagram by an analytical COMF approach. The results
show exact agreement between numerical simulations and
our theory.

The analysis of the dynamical evolution of our new model 
may give us a clearer physical picture on
how the acceleration and stochastic delay rules affect the evolution 
and the corresponding asymptotic steady state. 
It will also provide us with 
better ideas on developing analytical 
approaches to other traffic flow CA
models such as the NS model for which no exact analytic approach 
has been established.

\acknowledgments
This work was supported in part by grants from the Hong
Kong Research Grants Council (RGC) and the Hong Kong 
Baptist University Faculty
Research Grant (FRG).  
BHW acknowledges the support from
a Special Fund for a Major 
National Basic Research Project (Project 973) 
in China, the National Basic Research Climbing-Up Project
 ``Nonlinear Science'' and 
the National Natural Science Foundation in China under Key Project 
Grant No.19932020 and Project Grant Nos.19974039 and 59876039. 
In CUHK, BHW was supported by the RGC Grant CUHK 4191/97P. 
We thank Dr. P.M.Hui for useful discussions and
critical reading of the manuscript.

%\newpage
\begin{center}
{\bf Figure Captions}
\end{center}

Figure 1: The fundamental diagram 
with the maximum car velocity $M=1$ and for different 
stochastic delay probabilities $f$.  
The solid curves are the theoretical
results. The points with different symbols represent results of
numerical simulations.  The curves from the top down 
along the velocity axis correspond to different  $f$ 
ranging from $f=0$ to $f=1$ in steps of $0.1$.

\vspace*{0.15 true in}

Figure 2: The fundamental diagram  
with the maximum car velocity $M=2$ and for different 
stochastic delay probabilities $f$.  
The solid curves are the theoretical
results. The points with different symbols represent
numerical simulations.  The curves from the top down 
along the velocity axis correspond to different $f$ 
ranging from $f=0$ to $f=1$ in steps of  $0.1$.

\end{document}